# Extracting seizure onset from surface EEG with Independent Component Analysis: insights from simultaneous scalp and intracerebral EEG

Andrei Barborica[a,†,*], Ioana Mindruta[b,c,†], Laurent Sheybani[d], Laurent Spinelli[d], Irina Oane[b], Constantin Pistol[a], Cristian Donos[a], Víctor J López-Madrona[e], Serge Vulliemoz[d], Christian-George Bénar[e,*]

[a] Physics Department, University of Bucharest, 405 Atomistilor Street, Bucharest, Romania

[b] Epilepsy Monitoring Unit, Neurology Department, Emergency University Hospital Bucharest, 169 Splaiul Independentei Street, Bucharest, Romania

[c] Neurology Department, Medical Faculty, Carol Davila University of Medicine and Pharmacy Bucharest, 8 Eroii Sanitari Blvd 8, Bucharest, Romania

[d] EEG and Epilepsy Unit, University Hospitals and Faculty of Medicine Geneva, Rue Gabrielle-Perret-Gentil 4, 1205 Geneva, Switzerland

[e] Aix Marseille Univ, INSERM, INS, Inst Neurosci Syst, Marseille, France

E-mail addresses: andrei.barborica@fizica.unibuc.ro (A. Barborica), ioanamindruta@me.com (I. Mindruta), laurent.sheybani@hcuge.ch (L. Sheybani), laurent.spinelli@hcuge.ch (L. Spinelli), dr.popairina@gmail.com (I. Oane), costi.pistol@gmail.com (C. Pistol), cristian.donos@g.unibuc.ro (C. Donos), victor.lopez-madrona@univ-amu.fr (V.J. López-Madrona), serge.vulliemoz@hcuge.ch (S. Vulliemoz), christian.benar@univ-amu.fr (C.G. Bénar)

[†] The authors have equally contributed to this work

[*] Corresponding authors:
Christian-George Bénar
christian.benar@univ-amu.fr
Aix Marseille University, Inserm, INS, Institut de Neurosciences des Systèmes, INS, UMR 1106, Faculté de Médecine La Timone, 27 Bd Jean Moulin Marseille F-13005, France

Andrei Barborica
andrei.barborica@fizica.unibuc.ro
Physics Department, University of Bucharest, 405 Atomistilor Street, Bucharest 077125, Romania

**Abstract**

The success of stereoelectroencephalographic (SEEG) investigations depends crucially on the hypotheses on the putative location of the seizure onset zone. This information is derived from non-invasive data either based on visual analysis or advanced source localization algorithms. While source localization applied to interictal spikes recorded on scalp is the classical method, it does not provide unequivocal information regarding the seizure onset zone. Raw ictal activity contains a mixture of signals originating from several regions of the brain as well as EMG artifacts, hampering direct input to the source localization algorithms. We therefore introduce a methodology that disentangles the various sources contributing to the scalp ictal activity using independent component analysis and uses equivalent current dipole localization as putative locus of ictal sources. We validated the results of our analysis pipeline by performing long-term simultaneous scalp – intracerebral (SEEG) recordings in 14 patients and analyzing the wavelet coherence between the independent component encoding the ictal discharge and the SEEG signals in 8 patients passing the inclusion criteria. Our results show that invasively recorded ictal onset patterns, including low-voltage fast activity, can be captured by the independent component analysis of scalp EEG. The visibility of the ictal activity strongly depends on the depth of the sources. The equivalent current dipole localization can point to the seizure onset zone (SOZ) with an accuracy that can be as high as 10mm for superficially located sources, that gradually decreases for deeper seizure generators, averaging at 47 mm in the 8 analyzed patients. Independent component analysis is therefore shown to have a promising SOZ localizing value, indicating whether the seizure onset zone is neocortical, and its approximate location, or located in mesial structures. That may contribute to a better crafting of the hypotheses used as basis of the stereo-EEG implantations.

## 1. Introduction

In the context of epilepsy presurgical evaluation, intracerebral EEG is a reference method for defining the epileptogenic zone (EZ) that needs to be resected in order to render the patient seizure free (Bartolomei et al., 2017; Kahane et al., 2006; Talairach and Bancaud, 1973). It possesses exquisite spatial specificity and signal-to-noise ratio across a large frequency band (Roehri et al., 2016), but only a limited number of electrodes can be implanted. Thus, the success of the invasive phase depends crucially on the hypotheses on the putative EZ location that are obtained from non-invasive data. Among non-invasive techniques, electroencephalography (EEG) remains the tool of choice for long term monitoring. Typically, only a low number of electrodes are used during long term monitoring (between 19-32), rendering source localization difficult. Even with a higher number of electrodes, the fast discharges that are usually present at seizure onset can be difficult to localize (Koessler et al., 2010) Moreover, movement artefacts and in general difficulty of preserving signal quality over the long term contribute to making ictal electric source analysis a challenging technique (van Mierlo et al., 2020).

Blind source separation such as independent component analysis (ICA) (Comon, 1994) has proven useful for separating independent sources linearly mixed in the recorded EEG signals, including artifacts. One of the primary uses of ICA was to perform artifact subtraction in EEG signals (Jung et al., 2000). In the context of epilepsy, Canonical Correlation Analysis has been proposed in order to denoise scalp EEG traces during seizures by removing components that present little auto-correlation (De Vos et al., 2007; Vergult et al., 2007). In our approach, we hypothesize that the ictal onset signal can be identified as an independent source in the EEG, in addition to the physiological and non-physiological signals and artefacts. This has already been attempted in MEG for epilepsy (Malinowska et al., 2014; Ossadtchi et al., 2004; Pizzo et al., 2019). The localization of the ictal sources, disentangled by ICA from background brain activity or noise, could thus provide valuable information regarding the location of the seizure onset zone (SOZ) and seizure spread areas.

Intracerebral EEG during presurgical evaluation of epilepsy offers the unique opportunity to compare non-invasive results to the activity recorded directly within the brain. While there is a significant amount of studies performing a comparative analysis of sequentially recorded scalp and invasive activity, the literature dedicated to simultaneous scalp and SEEG, and in particular to (pre)ictal activity is scarce. It has been shown recently that it is possible to record simultaneously surface and depth measurements, thus obtained different views on the exact same brain activity (Abramovici et al., 2018; Antony et al., 2019a; Koessler et al., 2015, 2010; Ramantani et al., 2016).

In this study, we used simultaneous long-term EEG-SEEG recordings in 14 patients during their presurgical evaluation of epilepsy, and high-density EEG-SEEG recording in one patient, as a proof-of-concept for the use of ICA for recovering epilepsy-related activity at seizure onset.

## 2. Methods

### 2.1. Patient selection

We selected 14 consecutive patients diagnosed with focal drug resistant epilepsy that underwent long-term simultaneous EEG and stereoelectroencephalographic (SEEG) recordings in the Emergency University Hospital Bucharest between 2018 and 2020 (Table 1). Patients were considered surgical candidates and underwent presurgical non-invasive evaluation using extended patient history, video-electroencephalography, brain structural and functional imaging (inter-ictal FDG-PET CT) and neuropsychological profile. Consequently, in these patients, invasive recordings were considered necessary to delineate the epileptogenic zone and map functional cortex to tailor the surgical resection (Isnard et al., 2018; Jayakar et al., 2016; Kahane et al., 2003; Munari et al., 1994). The details regarding the patients' gender, age, type of epilepsy and lateralization are provided in Table 1. In addition, part of this research protocol, scalp electrodes were attached, allowing for simultaneous surface and intracranial long-term recordings.

We excluded from further analysis 6 patients in which: i) we were not able to record a typical seizure during simultaneous scalp-SEEG monitoring (*n*=4), ii) there were no typical ictal discharges visible on scalp at seizure onset (*n*=1), ii) the seizure onset zone was not focal (*n*=1). The eight remaining patients recorded in the Bucharest center were kept for further group analysis.

The study has been performed under Bucharest University ethical committee approval CEC 23/20.04.2019. All patients signed a written informed consent, in accordance with the Declaration of Helsinki, for the simultaneous recordings and data sharing procedures.

*2.2. Long-term EEG-SEEG recordings*

SEEG exploration was performed using depth electrodes (Dixi Medical, Chaudefontaine, France) with 8 to 18 contacts per electrode, 2 mm contact length, 3.5 mm center-to-center contact spacing and 0.8 mm diameter. Multiple electrodes were placed following an individual hypothesis allowing for up to 258 contacts to be available in each patient. Electrodes were placed intracranially using the Leksell stereotactic frame (Elekta AB, Stockholm, Sweden) or the microTargeting™ Multi-Oblique Epilepsy STarFix Platform (FHC, Bowdoin, ME USA) (Dewan et al., 2018; Pistol et al., 2020; Yu et al., 2018). To determine the exact location of each electrode and contact, the post-implantation CT scan was loaded in the surgical planning software (Waypoint Planner, FHC, Bowdoin, ME USA), co-registered with the pre-implantation MRI, and adjustments to the initially planned trajectories were made to match the postop location of the electrodes.

Between 20 and 37 scalp electrodes were placed according to the 10-20 system. A few electrodes were repositioned on adjacent 10-10 grid location, due to interference with the SEEG electrodes and up to 10 electrodes could not be placed at all. The exact number of scalp electrodes in each patient is provided in Table 1.

To ensure maximal scalp and intracranial signal quality and comparability, two identical Natus Quantum 128-channel amplifiers (Natus Neuro, Middleton, WI) were used, galvanically isolated and having separate hardware references. The hardware reference for the SEEG recordings was chosen on one contact located in white matter exhibiting minimal iEEG activity, whereas for the scalp system the reference was Fpz. The use of an intracranial reference for the SEEG recordings guarantees that the SEEG signals are not contaminated by EMG artefacts during ictal events. Raw data was acquired using the same sampling rate of 4096 Hz for both systems. The use of two independent systems posed the challenge of continuously synchronizing the recordings for the entire duration of the investigation. This has been performed using a synchronization hardware that sent periodically digital triggers to both systems every hour. To fine-tune the drift in the clocks of the two systems in the interval between two trigger pulses, a 50 Hz sine reference signal, derived from the power lines using a step-down isolation transformer was applied simultaneously to DC inputs of the two systems. Data segments around ictal events were loaded from scalp EEG and SEEG data files, aligned based on the synchronization digital trigger channel. Additionally, a cross-correlation between the reference sine waves recorded on each channel was used to calculate the lag between the recordings at any time and make the appropriate corrections to perfectly align the recordings. The data was combined and saved in a single file in AnyWave ADES format (Colombet et al., 2015), containing both types of signals. The correct operation of the synchronization hardware and software was validated using single-pulse electrical stimulation, now part of our routine clinical investigation protocol (Donos et al., 2016a, 2016b), by verifying that the stimulation artifacts, visible on both scalp and SEEG recordings are perfectly aligned.

Video-EEG-SEEG recordings were performed in chronic conditions for 7 to 14 days at the Emergency University Hospital Bucharest. The scalp electrodes were attached 1 to 3 days after the SEEG recording started. While maximum care was paid to ensure lowest electrode impedance during the initial positioning, the quality of the scalp signal degraded over time, making the scalp recordings usable for a period of up to 4 days. At the end of this interval, the scalp electrodes were removed and intracranial recording continued, as needed, in order to complete the rest of the recording and stimulation protocols part of the presurgical evaluation.

*2.3. Data Selection*

Ictal epochs of 30 seconds were selected by visual inspection of the SEEG traces, based on a marker of seizure onset (SO) placed manually by the expert epileptologist. The section included 20 seconds before SO (incorporating the preictal state) and 10 seconds after. For subsequent visualization and analysis, intracranial data was re-referenced using a bipolar montage. The selected intervals were exported to Brainvision data format (Brain Products GmbH, Munich, Germany) thanks to our in-house software AnyWave (Colombet et al., 2015) (available at meg.univ-amu.fr) or read directly from AnyWave's ADES files. All further data processing was performed with the EEGLAB toolbox (Delorme and Makeig, 2004) and custom Matlab (Mathworks, Natick, MA) code.

*2.4. ICA and dipole localization*

Scalp EEG signals were filtered in the 5 Hz-100 Hz band using the *eegfiltnew* function of EEGLAB (Hamming windowed sinc FIR filter, with default transition bandwidth of 2 Hz in this case). Then infomax ICA (Bell and Sejnowski, 1995) was performed on these signals (*runica* function, with 'extended' option). A equivalent current dipole (ECD) model was fitted to each component in a standard boundary element model (BEM) volume in MNI coordinates. Using patient-specific head models based on the patient's own MRI was considered as having a limited benefit due to the imprecision arising from the fact that the EEG electrode location was not digitized. We automatically removed components with at least one outlier electrode (Z-value across electrodes >10). Only components presenting a dipolar topography were retained (goodness of fit GOF>90%), considered as putative brain sources (Malinowska et al., 2014). The independent components best encoding the ictal activity were then manually selected by an expert epileptologist among candidates that were provided initially by the analysis software based solely on the goodness of fit value (Sohrabpour et al., 2020).

We also performed distributed sources analysis, sLoreta (Loreta-Key software, Pascual-Marqui et al., 1994) and beamformer linearly constrained minimum variance (LCMV) using MNE-Python package (Gramfort et al., 2014) for the selected independent components. For Loreta, we used the relative regularization technique, with a value of 1.

*2.5. Coherence*

We computed wavelet coherence (Lachaux et al., 2002) between each ICA component and each SEEG signal, based on Morlet wavelets with oscillation parameter ξ=7 (Bénar et al., 2009). We computed the time-frequency analysis in log scale, between 5 Hz and 100 Hz, with 10 voices per octave. For visualization purposes of each time frequency plane, but not for subsequent analysis, a Z-score normalization was applied to the plotted time-frequency decompositions (Roehri et al., 2016). In order to compute the mean and standard deviation used in the Z-score, we used only the points in lowest amplitude (first 20% quantile). For computing coherence, the time-frequency representation of one signal was multiplied by the conjugate of the other, and the resulting image was smoothed in time and frequency with a rectangular kernel (Grinsted et al., 2004). The width of a wavelet at a given frequency $f$ was computed as $2\sigma = \xi/(\pi f)$, and the extent of the smoothing window was taken empirically as 10 times the width of a wavelet (i.e., $20\sigma$) along time and 10 voices along frequencies.

*2.6. Statistics*

We first estimated a threshold of significance at *p*=0.05 for each frequency based on random noise. We performed 200 realizations of white noise comprising 2 channels, and computed coherence between these channels with the same parameters as real data (signal length, $\xi$ and smoothing parameters). We verified in simulations that changing the slope of the noise (-pink instead of white) did not lead to major changes in threshold (not shown). We also tried on one pair ICA/SEEG using using surrogate data based on phase scrambling. This lead to slightly higher sensitivity (not shown), but was not chosen because of the computational burden. These thresholds were applied to coherence measured on real data and counted for each coherence time-frequency plane (corresponding to a given pair of signals) the number of significant points *Ns* in the time-frequency plane. The quantification of the level of coherence in a given time-frequency plane was done by dividing the number of points *Ns* by the total number of points (excluding edge effects), resulting in a proportion of significant coherent points or significant normalized area (SNA). We then detected the outliers in the distribution of *Ns* across all pairs. The threshold was set as $m+5*(Q_{0.75}-Q_{0.25})$, with *m* median of *Ns* and $Q_{0.75}$, $Q_{0.25}$ quartiles of the distribution. This permitted to select the pairs of channels presenting the highest number of significant points.

*2.7. 3D Spatial Analysis and Visualization*

Patients' MRIs were registered to the FreeSurfer's *cvs_avg35* template in MNI space using a combined volumetric and surface (CVS) registration method (Postelnicu et al., 2009) available in FreeSurfer software package (Fischl, 2012) (available at http://surfer.nmr.mgh.harvard.edu). Unlike the linear registration methods that rely on transformation matrices, this registration method creates a warp field that maps a voxel in the subject space to one or more voxels in the template space. Therefore, to warp the contact coordinates from subject space to the MNI space we created volume masks of contacts, we warped the masks, then we assigned the contact the MNI coordinates of the new masks' average coordinates of the non-zero voxels.

Using the method described above, the actual location of the SEEG electrodes in patient's MRI coordinates were converted to MNI coordinates and compared to the dipoles' MNI coordinates for all patients (Postelnicu et al., 2009).

The Euclidean distances between scalp electrodes, SEEG contacts and dipoles corresponding to the independent components for the scalp EEG were computed in MNI space. Distance profiles between dipoles and SEEG contacts were created along with coherence profiles between SEEG and independent components. The SEEG contacts that are part of the seizure onset zone (SOZ) were identified by an expert epileptologist based on a typical seizure onset pattern as described in Perucca et al., (2014) and follow-up after the resective surgery. The 3D geometrical center of SOZ was computed by averaging MNI coordinates of SOZ contacts. The spatial relationship between SOZ and dipole locations was analyzed, and average SOZ channels' coherence with the independent components was displayed for review.

*2.8. Data availability*

Thirty seconds (up to 300 s) of simultaneous scalp-SEEG recordings at seizure onset in AnyWave-compatible format, as well as SEEG electrode locations in MNI coordinates can be found at

http://epi.fizica.unibuc.ro/scalesictal/. The additional source code and data can be provided by the authors upon reasonable request.

**3. Results**

*3.1. Single-patient example 1*

The results for patient 3 are shown in Figures 1 and 2. The SEEG implantation of the electrodes is shown in 3D (Fig. 1A), coronal (Fig. 1B) and sagittal (Fig. 1C) views. The combined scalp – intracranial EEG signals of a habitual seizure are shown in Fig. 1 D. The activity on V07-V08 pair, located in the inferior parietal lobule, exhibited a sequence consisting in repetitive low-frequency spikes, followed by a burst of polyspikes and a higher frequency ~10 Hz repetitive discharge (Perucca et al., 2014) and was considered to be the seizure onset zone. The independent component 4 (Fig. 1F,H) of the scalp recordings best captured the higher amplitude low-frequency spiking activity and exhibited a significant coherence (Fig. 1I) with the signal recorded from SOZ (Fig. 1E,G). Based on the component 4 topography, we also checked the localizing value of the scalp sensors alone by performing a coherence analysis between P4 and intracranial electrodes. The results are presented in Supplementary Figure S1 and indicate a weaker and less specific coherence with SEEG electrodes.

The 2D top view of the component field showing also the locations of the scalp electrodes is shown in Fig. 2A, whereas the 3D spatial location of the dipole #4 relative to the SEEG electrode in MNI space is shown in Fig. 2C-E. The coherence between the component #4 and the activity on all SEEG electrodes is shown in Fig. 2B. It is also indicated in Figs. 2C-E through the use of larger and brighter markers for the SEEG contacts exhibiting higher coherence with the selected component. The Euclidean distance from the dipole to each SEEG contact pair is shown using red circles in Fig. 2B. It can be seen that the maximum coherence between the component and the iEEG signals is for the V07-V08 pair, located 38.6 mm from the dipole, that is part of the seizure onset zone (SOZ). However, the dipole is located more mesial, with the nearest SEEG electrode pair being the tip of the V electrode, V01-V02, at 3.5 mm. Considering all the SEEG pairs, the correlation between coherence and distance is negative, $r$=-0.39 (Pearson, $p$<0.001).

*3.2. Single-patient example 2*

The second patient (7) used to illustrate the results of the analysis pipeline had a SOZ located superficially in the angular gyrus, covered by contacts V07 through V16 (Fig. 3A-C). The ictal discharge pattern consisted in low-frequency periodic spikes followed by low-voltage fast activity (LVFA) (Fig. 3 D-E). The analysis interval used the LVFA onset as reference. Supplementary figures S2 and S3 show the time-frequency (TF) maps of the scalp signals, showing that the LVFA pattern is barely visible on scalp (electrode P4, also C4 and T8), facilitated primarily by the Z-score TF normalization (Roehri et al., 2016). Overcoming this poor visibility, the independent component analysis is able to identify a component (IC10) that clearly encodes the ictal activity (Fig. 3H). The time-frequency maps of the signal recorded in SOZ (V'16-V'17) and of the independent component 10 from the analysis appear to be equally capturing the LVFA component, which is confirmed by the results of the wavelet coherence analysis as shown in Fig. 3I.

The dipole corresponding to component #10 was located close to the SOZ (nearest SOZ contact is V12, $d$=4.4 mmm), as it can be also seen in Fig. 4C-E, red circles. As a note specific to this patient, the patient had undergone a previous temporal lobe resection, therefore the dipole fitting, CVS coregistration and calculation of the MNI electrode coordinates may have been affected by the modified patient anatomy.

Just like for the previous patient, we have also checked the localizing value of the signal on the P4 scalp electrode, results presented in Supplementary Figure S4, showing a similar coherence profile as IC10, but with lower overall magnitude of the coherence.

*3.3. Frequency-dependent scalp visibility of intracranial activity*

Seizures recorded in patient 12 represents an illustrative example of a high-frequency discharge (~60–80 Hz) present simultaneously with a low-frequency (~4 Hz) repetitive discharge on a contact pair located in the orbitofrontal cortex. The scalp visibility of the two spectral components is different, the high-frequency discharge being invisible, whereas the low-frequency discharge is captured by independent component 3, as illustrated in Supplementary Figure S5.

*3.4. Group analysis*

Group analysis was performed for eight patients with long-term recordings, where we have recorded spontaneous seizures, after excluding six patients not having a seizure for the duration of the monitoring (n=4, patients 1, 6, 8, 10), one patient where scalp activity was affected by early EMG artifacts (patient 4), and one with non-focal epilepsy (patient 11).

For the entire patient cohort included in the analysis ($n$=8), we have investigated the relationship between the representative ICs and SEEG pairs exhibiting maximum coherence with ICs, as well as the SOZ localizing value of the ICs (Fig. 5). While the two illustrative patients had rather superficially located SOZs, the remaining 6 had deeper SOZs located in the insula ($n$=3), amygdala ($n$=1), medial orbitofrontal cortex ($n$=1) and posterior peri-ventricular ($n$=1). Fig. 5A presents the distance between dipoles and the SEEG contact pairs exhibiting the highest coherence, $d_{mean}$ = 60.4 ± 32.5 mm (mean ± SD). Fig. 5B illustrates a monotonic negative relationship between maximum coherence and distance to equivalent current dipole, pointing to the fact that high coherence is associated with a low localization error. The Spearman's correlation coefficient for the raw data points in Fig. 5B is $\rho$=-0.79 ($p$<0.05), however these results should be cautiously regarded due to small sample size. Regarding SOZ, Fig. 5C shows the mean value of the coherence between selected independent component (following criteria described in the methods section) and SEEG signals recorded from SOZ, mean value across patients $SNA$ = 0.059 ± 0.078. It can be seen that the coherence between SEEG and scalp components is high for the superficially located SOZ (patients 3, 7) and much smaller for deeper SOZ (patients 2, 5, 9, 12, 13, 14). The SOZ localization value of the dipoles associated with the representative independent components is shown in Fig. 5D, where we have plotted the Euclidean distance between geometrical center of SOZ and dipoles' location, $d_{SOZ}$ = 47.2 ± 23.2 mm. The mean coherence of the SOZ contacts with the ICs is shown in Fig. 5E-G using markers whose size is proportional to the coherence, and the marker location corresponds to SOZ, by using the mean of the 3D coordinates of the contacts located in SOZ. We have cross-checked the equivalent current dipole approach against distributed source localization methods (sLoreta, Beamformer LCMV) and presented the results in Supplementary Figure S6. None on the distributed source localization algorithms provided results that are better, on average, than ECD. The mean distance between peak activation and geometrical center of SOZ was $d_{SOZ}$ = 49.7 ± 26.1 mm for sLoreta and $d_{SOZ}$ = 55.1 ± 25.4 mm for beamformer LCMV.

## 4. Discussion

There are few validation studies of visibility of ictal activity on scalp EEG (Koessler et al., 2010). Previous studies using simultaneous scalp EEG-SEEG recordings have primarily focused on the detectability of the interictal spikes on scalp recordings (Koessler et al., 2015), but also on the resting-state activity (Fahimi Hnazaee et al., 2020). Recently, the ictal activity that contains repetitive and high-frequency components has been investigated as well (Abramovici et al., 2018; Antony et al., 2019a, 2019b). However, in these studies, the correspondence between scalp and intracranial activity has been based exclusively on a visual analysis. A recent study by Sohrabpour et al., (2020) implements an approach based on independent component analysis of ictal activity and distributed source imaging, however it does not benefit from simultaneous scalp-iEEG recordings. Our study implements an entirely quantitative analysis that characterizes the coherence between intracranial and scalp activity, as well as the single-source localization of the dipoles associated with independent components encoding ictal activity, providing the best guess for the SOZ localization.

While recent reports indicate that deep subcortical activity (originating in the nucleus accumbens and centromedial thalamus) is detectable on scalp recordings (Seeber et al., 2019), our results show that the visibility of the ictal discharges strongly depends on how deep the SOZ is located. Factors behind this difference may include the frequency range and the amplitude of the ictal signals, which are significantly different from the slow activity (<10 Hz) being considered for subcortical structures. A study by Tao et al. (2007) showed that only cortical ictal patterns with low frequency, between 2 to 9 Hz, were visible on scalp. The same study indicated that a minimum cortical source area exhibiting synchronous activity of >10 $cm^2$ is mandatory for generating scalp-recordable ictal EEG patterns. As a result of these concurring constraints, less than half (40%) of the seizures in our study could be associated with a nearly simultaneous pattern on both scalp and iEEG. Many of the LVFAs recorded during typical SEEG investigations have significantly higher characteristic frequencies and are highly focal, without entraining large cortical patches, at least early during the seizure development. This might explain why only a subset of our patients provided concordant scalp-SEEG recordings allowing for an accurate localization of the ictal source activity.

We have found that superficially located SOZs have better visibility on scalp, in line with the evidence brought by previous studies on interictal spikes (Koessler et al., 2015) and high-frequency oscillations (Avigdor et al., 2020). The data from patient 3 and particularly from patient 7, provided here as examples, shows that not only the high-amplitude low frequency activity is visible on scalp, but also the low-voltage fast activity of the order of ~60 Hz. Independent component analysis can identify a component (IC10, Fig. 3H) that encodes the ictal activity, highlighting the contribution of ICA to extract relevant information from surface EEG data. Similar LVFA patterns present in deeper structures (insular cortex, medial orbitofrontal cortex or peri-ventricular) were not captured by scalp activity, at least not with our proposed analysis pipeline. In Supplementary Figure S5 we show an example of a high-frequency discharge (~60–80 Hz present simultaneously with a low-frequency (~4 Hz) repetitive discharge on a contact pair deeply located in the orbitofrontal cortex of patient 12. The low-frequency component is visible on the scalp, being captured by independent component 3, whereas the high-frequency component is not visible, not being captured by any of the independent components. The initial focal LVFA is a trigger for more widespread repetitive discharges entraining larger cortical patches. This example reinforces the evidence in support of the fact that the frequency of the discharges is crucial for the visibility of the deep ictal sources.

While average SOZ localization error is relatively high ($d_{SOZ}$ = 47.2 mm), primarily as a result of deep ictal sources poorly visible on scalp in some patients, the mere fact that in a subset of patients the ictal discharges can be captured by ICA and visually detected by an expert epileptologist, pointing to the SOZ with sub-lobar resolution, bears clinical value. The hypothesis regarding SOZ localization following phase I investigation is crafted based on a multitude of factors, possibly including the results of our analysis, if there is indication they are robust, primarily based on the visual analysis of the time-frequency maps, component topography and the dipole's and goodness of fit.

These data show the benefit of long-term simultaneous recordings that are critical to capture seizures. Multivariate analysis allows denoising the data and capturing activity that is distributed over the electrodes. Here, ICA provides for each component a topography that can be localized using source analysis methods (Malinowska et al., 2014). In comparing localizing value of ICA vs scalp sensors, based on coherence with intracranial signals, the ICA provides larger and more specific coherence value (compare figures 2 and 4 against S1 and S4, respectively). The algorithm can be driven by preictal activity in the same location as the seizure onset zone, which could be the case in our example 1, but potentially also by the actual fast discharge, as suggested by our example 2. Future studies might investigate spectrum whitening (through Z-score normalization, for instance, Roehri et al., 2016), that could potentially increase detectability of high frequency activity. We used dipole localization for finding the source of IC topographies, which is a very rough approximation of otherwise distributed sources (Sohrabpour et al., 2020). In interpreting our results, one has to keep in mind that the single equivalent current dipole should be rather regarded as the geometrical center of a larger cortical dipole sheet whose spatial extent is unknown but expected to be broad, of the order of several square centimeters (>6.5 cm2, Nunez and Srinivasan, 2006). Also, dipole localization is likely not the most appropriate method for ictal localization that involve large brain areas (Sohrabpour et al., 2020). Indeed, a modelling study of Kobayashi et al. (2005) has pointed out that large activated cortical patches result in deeper dipoles, with a minor loss in GOF. This might be indeed the case for ictal discharges that spread across large areas in the 10-second interval considered in our analysis. To clarify whether these known limitations of the ECD modeling play a major role, we have applied two alternate distributed source localization methods (sLoreta, beamformer LCMV) to our data, that however did not provide better SOZ localization results (Fig. S4). Future investigation might consider additional or better refined source localization algorithms and the use of a larger number of scalp electrodes.

Another limitation related to the equivalent dipoles' localization of independent components is that the SEEG skull anchors prevented the scalp electrodes to be placed in a complete, uniform, 10-20 system layout, with some 10-10 extensions. Some of the electrodes could not be placed at all and after eliminating artifacted contacts, a relatively low number of electrodes placed on irregular grid were used for source reconstruction (n=20 – 37). Also, more accurate results are expected when using boundary element head models derived from patient's anatomy instead of the one derived from the MNI template, for which we opted since digitized electrode coordinates were not available and in order to be able to perform an inter-subject comparison as shown in Fig. 5E-G.

With a relatively small number of scalp electrodes in patients 1 through 14, the accuracy of dipole localization is expected to be limited. In an approach similar to the one by Mikulan et al. (2020), that uses electrical stimulation as ground truth for source localization, we have obtained preliminary data indicating an error of 18.9 ± 8.1 mm (mean ± SD) for 135 stimulations in 6 of our patients (unpublished data). This additional analysis sets the expectations regarding the accuracy of the ECD approach for our limited and non-uniform coverage with scalp electrode.

Also, one has to keep in mind that a spontaneous ictal discharge visibility on a particular intracranial electrode contact does not represent the ground truth for its localization, the focus of the discharge may be located in an area not covered with SEEG electrodes. 3D brain coverage with SEEG electrodes is non-uniform, being driven by the initial hypothesis and anatomical constraints and may be sparse (Jayakar et al., 2016; Kahane et al., 2003).

A difficulty pointed by our study is the fact that ictal activity in deep brain structures is subject to large localization errors. However, the mere fact that an ictal discharge has been captured or not in the independent components has a promising localizing value, indicating that the seizure onset zone is neocortical or in mesial structures.

## 5. Conclusions

Using simultaneous scalp – SEEG recordings of ictal onset patterns we have proven the concept of a fully quantified analysis based on independent component analysis of scalp EEG. The dipole location associated with the independent component encoding the ictal activity provides an objective, indication of possible seizure generators. To our knowledge, this is the first study demonstrating the visibility of ictal low-voltage fast activity onset pattern on scalp electrodes thanks to intracerebral EEG. However, it seems that our proposed method is more suited to superficially located ictal sources. Deeply located high-frequency ictal onset discharges are more difficult to be captured. Overall, our findings impact on hypothesis localization in planning invasive explorations for drug-resistant epilepsy.

**Declaration of competing interest**

Andrei Barborica PhD is also Vice-President and Chief Technological Officer of FHC Inc, the manufacturer of the stereotactic fixture used in Bucharest. The other authors have nothing to disclose in relation to this work.

**Funding**

Part of this work was funded by a FLAG ERA/HBP grant from Agence Nationale de la Recherche "SCALES" ANR-17-HBPR-0005, Unitatea Executiva Pentru Finantarea Invatamantului Superior a Cercetarii Dezvoltarii si Inovarii COFUND-FLAGERA II-SCALES. We also acknowledge funding from Swiss National Science Foundation grants 169198, 170873 and 192749 to Serge Vulliemoz and from the Geneva Faculty of Medicine to Laurent Sheybani.

**Appendix A. Supplementary data**

Supplementary data to this article can be found at Neuroimage online.

**TABLES AND FIGURES**

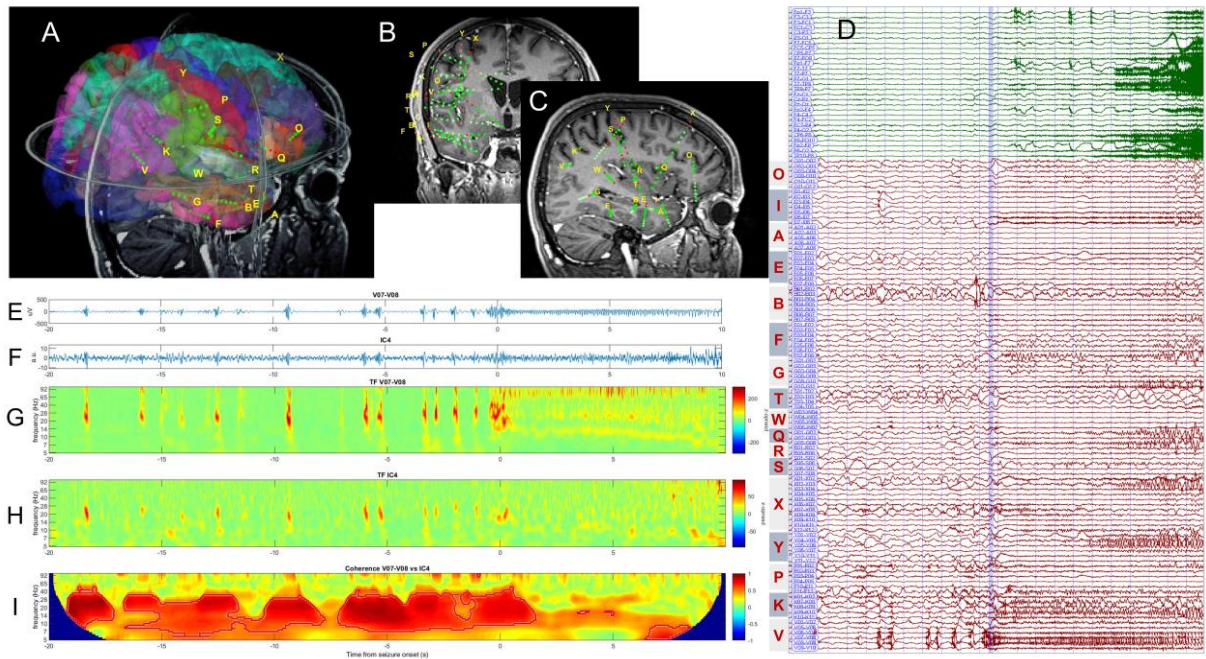

Figure 1. A-C) location of SEEG electrodes in patient 3 in 3D (A), coronal (B) and sagittal (C) views; D) combined scalp (green traces) and intracranial (red traces) recording of a seizure that originated from contacts $6-9$ of the electrode V implanted in the parietal cortex; E) intracranial EEG signal on pair V07-V08; F) Independent component 4 that best encoded the ictal and pre-ictal activity; G-H) time-frequency map of intracranial activity (depicted in E) and surface independent component (depicted in F); I) wavelet coherence between signals shown in E) and F).

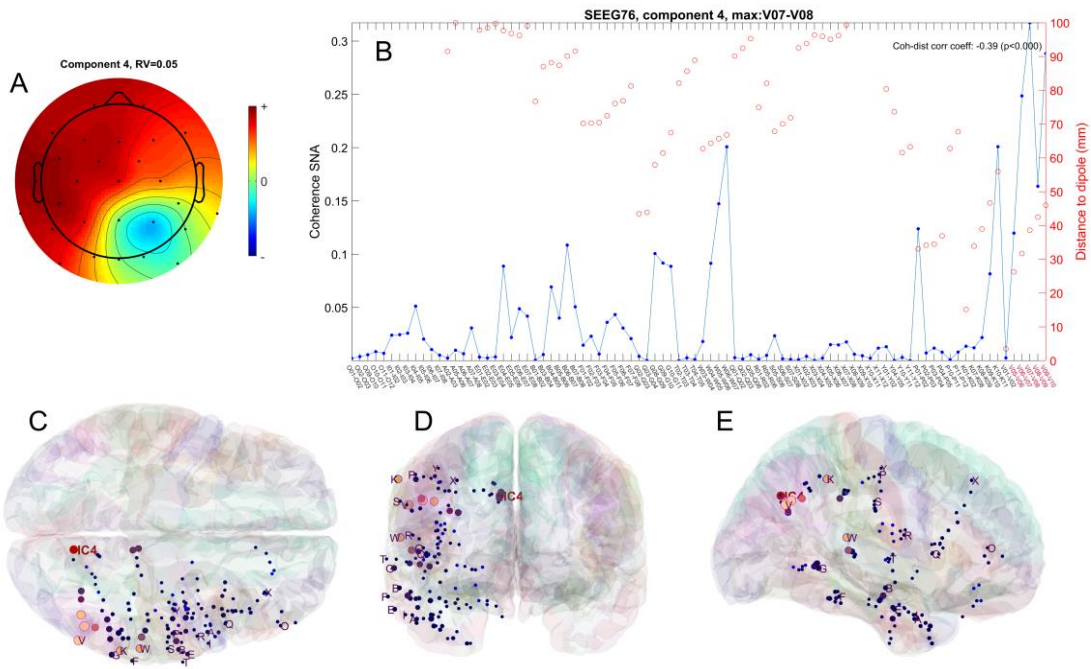

Figure 2. A) 2D topographic representation of the component 4 in patient 3, with the scalp electrodes location marked with black dots; B) coherence between component 4 and the intracranial signals (blue), along with the distance between dipole and contact pairs over which the signals were recorded (red); C-E) location of dipole corresponding to independent component 4 (red), as well as the SEEG electrodes (blue and pink), in MNI space, in axial (C), coronal (D) and sagittal (E) views; the size and brightness of the electrode markers is proportional to the coherence with the IC.

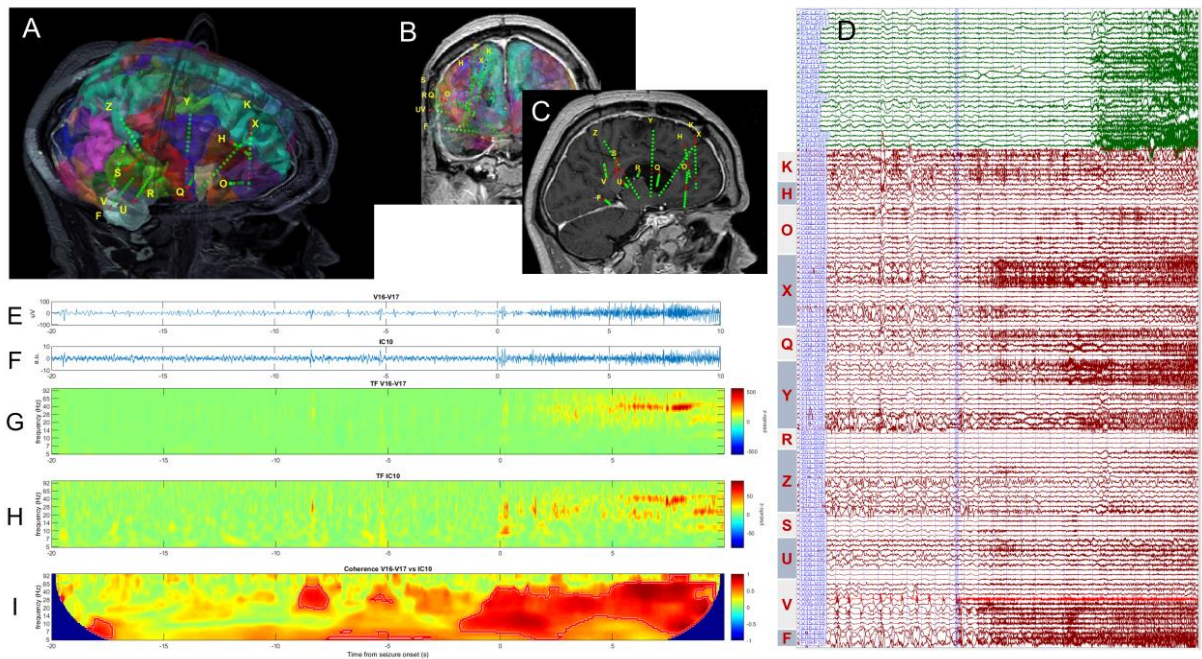

Figure 3. A-C) location of SEEG electrodes in patient 7 in 3D (A), coronal (B) and sagittal (C) views; D) combined scalp (green traces) and intracranial (red traces) recording of a seizure that originated from contacts 7 – 9 of the electrode V implanted in the angular gyrus; E) intracranial EEG signal on pair V16-V17; F) Independent component 10 that best encoded the ictal activity; G-H) time-frequency maps of the signal in E) and F, respectively; I) wavelet coherence between signals shown in E) and F).

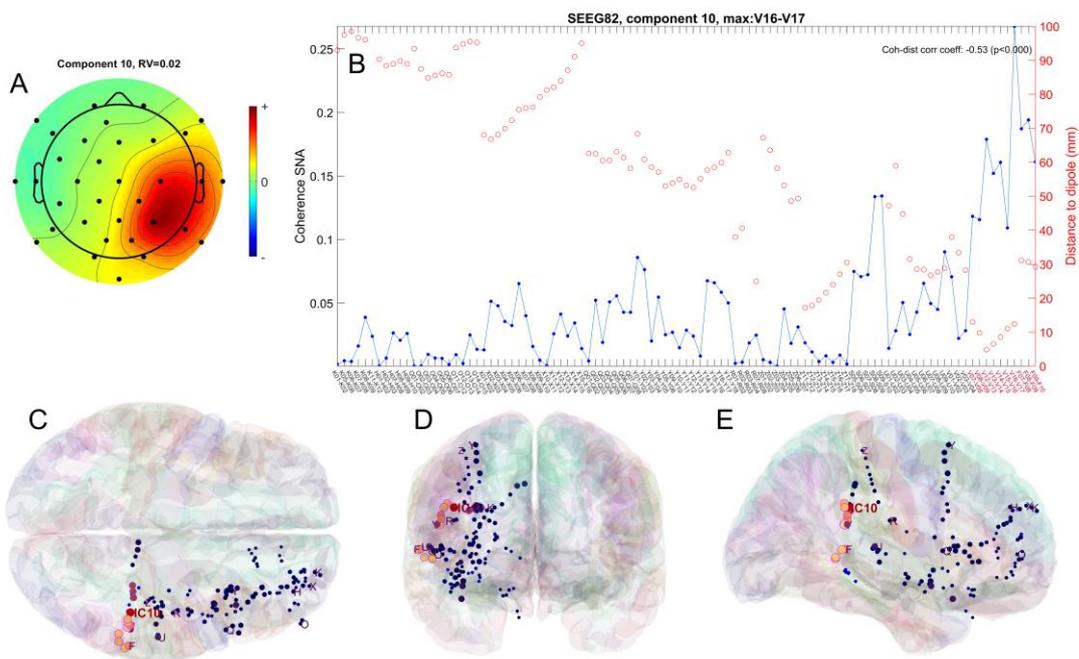

Figure 4. A) 2D topographic representation of the component 10 in patient 7, with the scalp electrodes location marked with black dots; B) coherence between the component and the intracranial signals (blue), along with the distance between dipole and contact pairs over which the signals were recorded (red); C-E) location of dipole corresponding to independent component 10 (red), as well as the SEEG electrodes (blue and pink), in MNI space, in axial (C), coronal (D) and sagittal (E) views; the size and brightness of the electrode markers is proportional to the coherence with the IC.

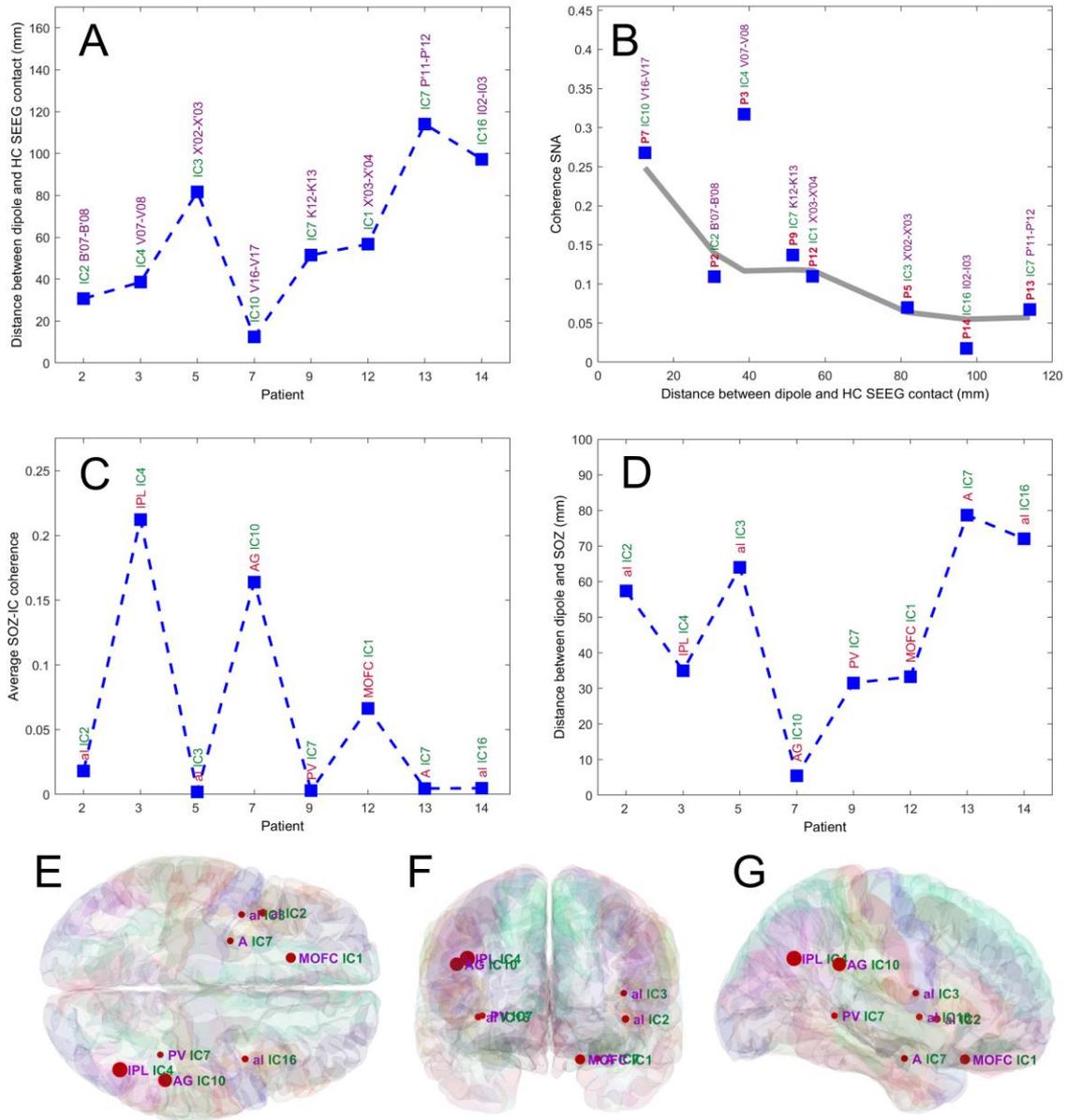

Figure 5. A) the distance in MNI space between component's dipole and SEEG pair exhibiting highest coherence (HC) with the component; the coherence is quantified as the significant normalized area (SNA) of time-frequency coherence analysis; B) magnitude of the HC as a function of the distance between HC SEEG pair and ECD; the gray line is the result of the locally weighted scatterplot smoothing (LOWESS); C) average coherence between selected independent components and signals on SEEG contacts located in SOZ; D) distance between component's dipole and SOZ location; E-G) 3D representation of the data in (A) using markers having a radius proportional to the coherence, at the actual SOZ location, in axial (E), coronal (F) and sagittal (G) views. Abbreviations used: A – amygdala, AG – angular gyrus, aI – anterior insula, IPL – inferior parietal lobule, PV – periventricular nodular heterotopia (in patient 9), MOFC – medial orbito-frontal cortex.

Table 1. Patients included in the analysis and characteristics of the ictal discharges.

| Patient | Sex | Age | Epi Onset Age | Lateralization | Epilepsy | SEEG Electrodes | SEEG Contacts | SEEG ictal pattern | Characteristic frequency (Hz) | Scalp Electrodes | Scalp ictal onset pattern | Scalp ictal spread pattern | Included |
|---|---|---|---|---|---|---|---|---|---|---|---|---|---|
| 1 | F | 29 | 27 | L | Temporal neocortical | 9 | 94 | No seizure | - | 20 | - | - | No |
| 2 | F | 24 | 9 | L | Insular | 14 | 163 | Bursts of polyspikes followed by LVFA | 70 | 21 | Not visible | Not visible | Yes |
| 3 | M | 28 | 26 | R | Parietal | 17 | 190 | Bursts of polyspikes followed by repetitive discharges | 10 | 28 | Visible | Visible | Yes |
| 4 | F | 19 | 2 | L | Insular | 13 | 155 | Baseline shift followed by LVFA | 40 | 23 | Not visible | Not visible | No |
| 5 | F | 17 | 13 | L | Insular | 12 | 168 | Bursts of polyspikes followed by LVFA | 50 | 24 | Not visible | Not visible | Yes |
| 6 | M | 19 | 12 | R | Temporal | 12 | 161 | No seizure | - | 32 | - | - | No |
| 7 | M | 40 | 6 | R | Parietal | 15 | 168 | LVFA at onset | 60 | 36 | Visible | Visible | Yes |
| 8 | F | 26 | 17 | R | Temporal-occipital | 18 | 258 | No seizure | - | 33 | - | - | No |
| 9 | M | 32 | 13 | R | Temporal-occipital | 15 | 214 | LVFA at onset | 30 | 33 | Not visible | Visible | Yes |
| 10 | M | 19 | 14 | R | Temporal-insular | 15 | 213 | No seizure | - | 35 | - | - | No |
| 11 | F | 26 | 17 | R | Frontal | 9 | 100 | LVFA at onset | 40 | 25 | Visible | Visible | No |
| 12 | F | 22 |  | L | Orbitofrontal | 15 | 191 | LVFA at onset | 60 | 37 | Not visible | Not visible | Yes |
| 13 | M | 47 | 46 | L | Temporal-mesial | 14 | 124 | LVFA at onset | 80 | 37 | Not visible | Not visible | Yes |
| 14 | M | 37 | 27 | R | Insular | 14 | 172 | LVFA at onset | 50 | 30 | Not visible | Visible | Yes |

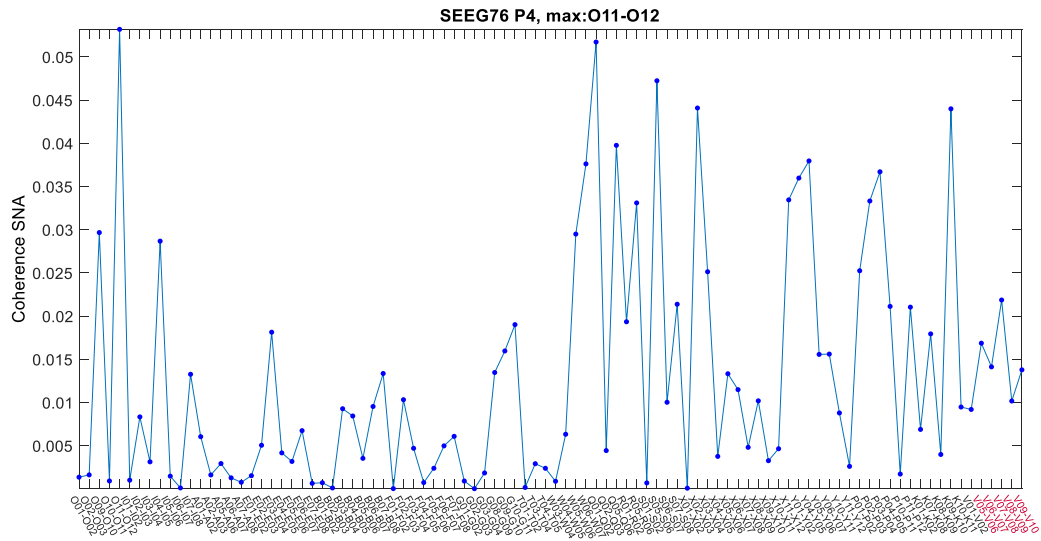

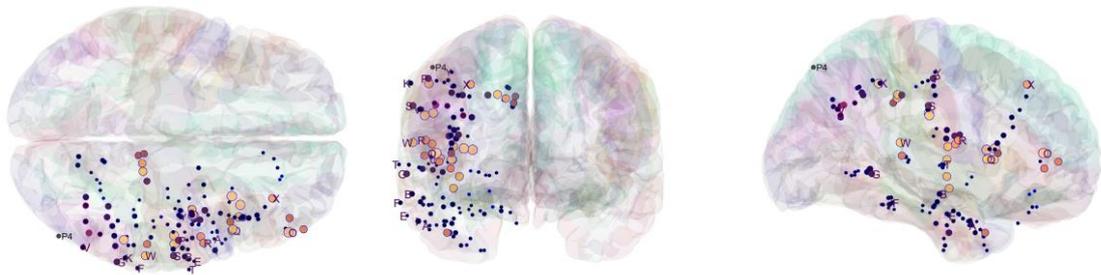

Supplementary Figure S1. Coherence of signal on scalp electrode P4 with intracranial electrodes in patient 3; A) top, profile of the coherence, quantified by the significant normalized area (SNA) in time-frequency plane; B-D) ) bottom row, 3D representation of SEEG electrodes (blue - pink), in MNI space, in axial, coronal and sagittal views; the size and brightness of the electrode markers is proportional to the coherence with P4.

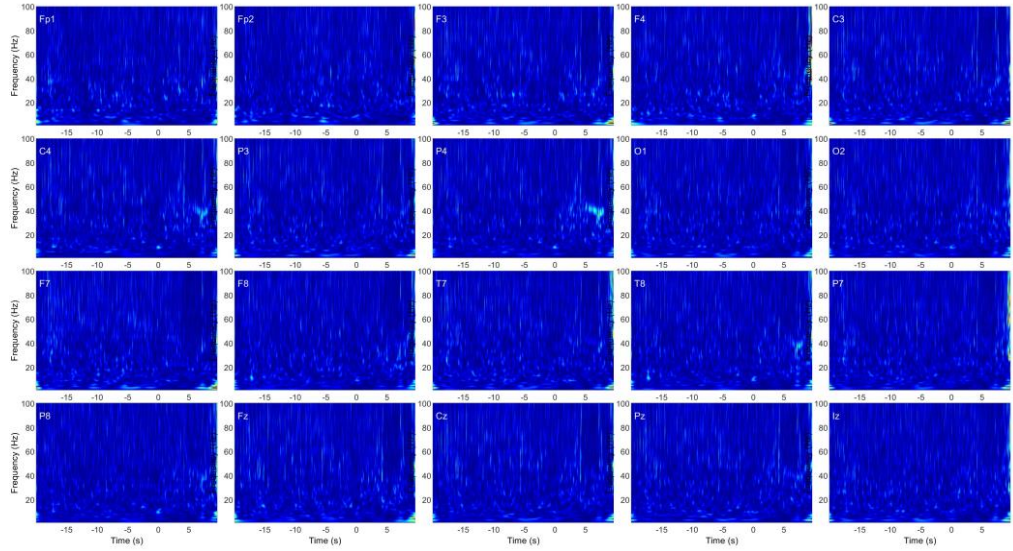

Supplementary Figure S2. Time-frequency maps (using Z-score normalization) of the scalp EEG recordings (first 20 out of 36) in patient 7 showing a ~40-Hz LVFA pattern visible on electrode P4.

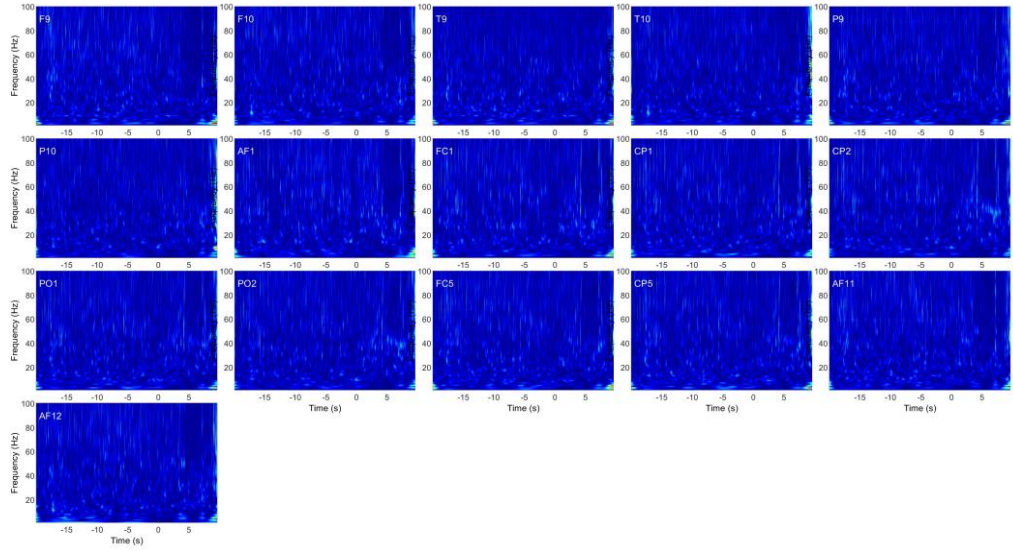

Supplementary Figure S3. Time-frequency maps (using Z-score normalization) of the scalp EEG recordings (channels 21 through 36) in patient 7, showing that LVFA patterns are not clearly visible on this signal set.

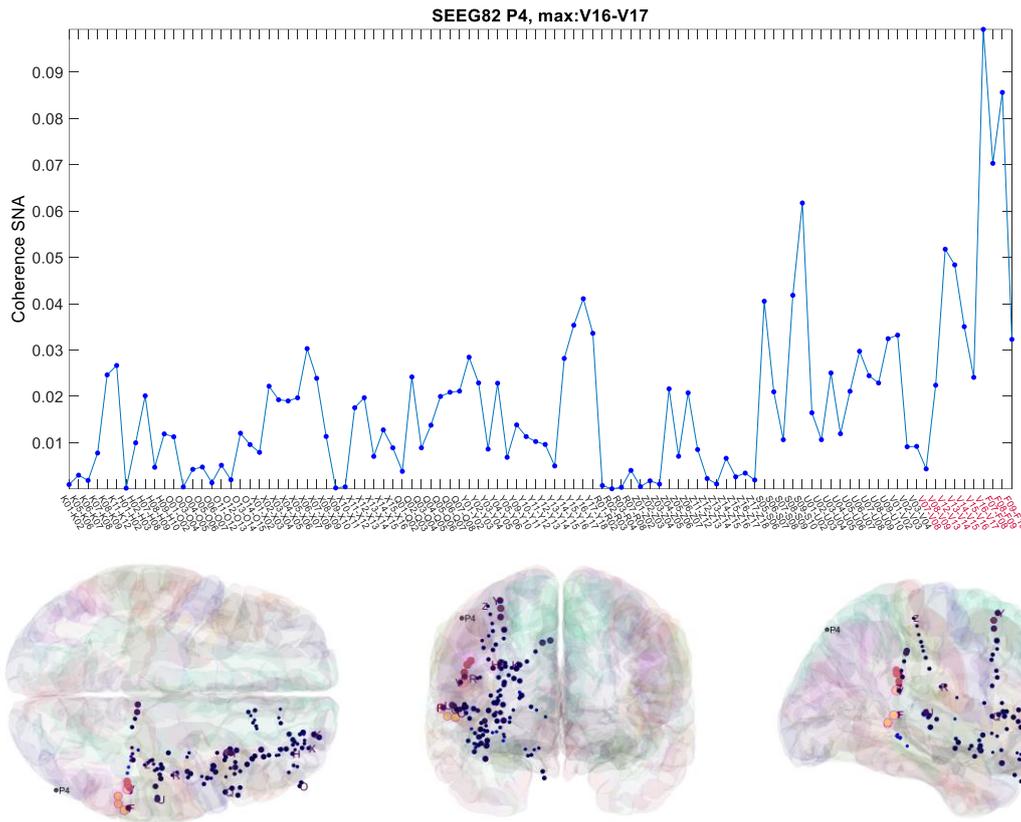

Supplementary Figure S4. Coherence of signal on scalp electrode P4 with intracranial electrodes in patient 7; A) top, profile of the coherence, quantified by the significant normalized area (SNA) in time-frequency plane; B-D) ) bottom row, 3D representation of SEEG electrodes (blue - pink), in MNI space, in axial, coronal and sagittal views; the size and brightness of the electrode markers is proportional to the coherence with P4.

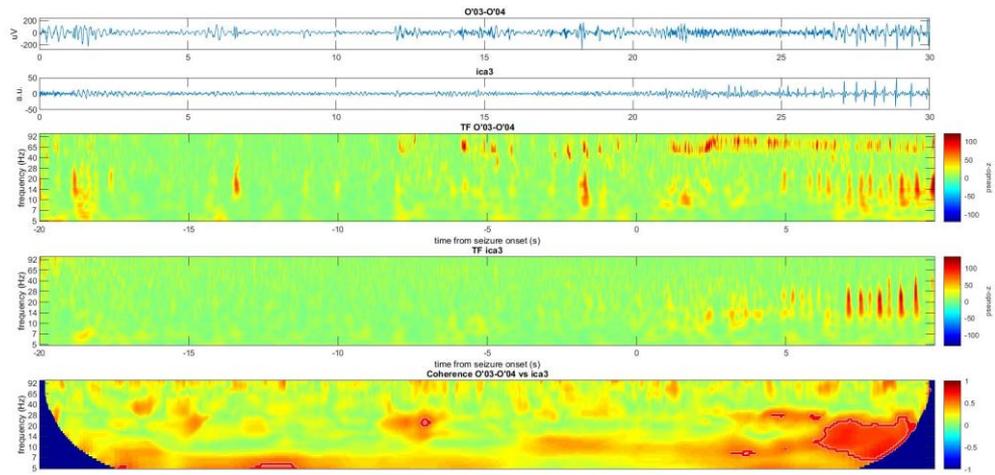
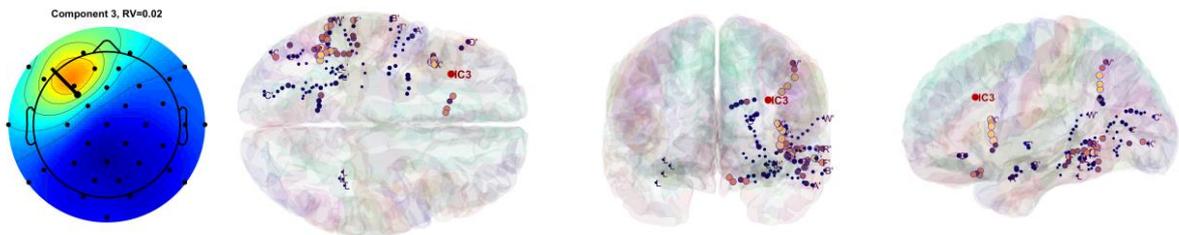

Supplementary Figure S5. Example of an ictal discharge in the SOZ of patient 12 (contacts O'03-O'04 located in orbitofrontal cortex) containing an initial LVFA followed by low-frequency repetitive discharges that propagate primarily to the insular cortex. Only low-frequency discharges originating from this deep source is visible on scalp and captured by the independent component 3. None of the other components capture LVFA.

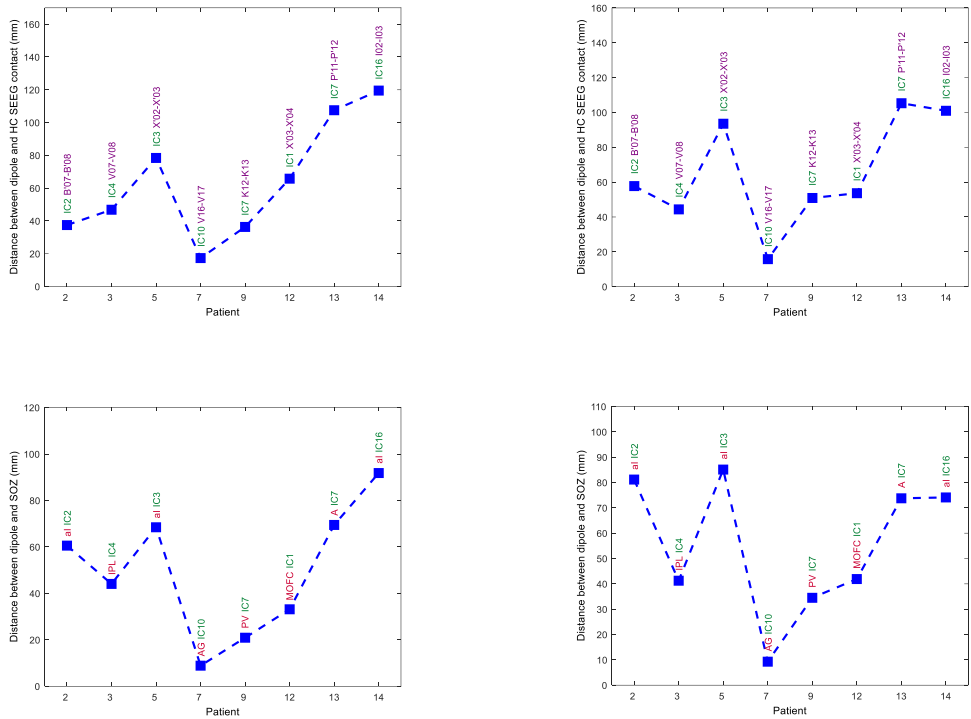

Supplementary Figure S6. Results of distributed source localization; A-B) the distance in MNI space between component's peak activation (A – sLoreta, B – beamformer LCMV) and SEEG pair exhibiting highest coherence (HC) with the component; C-D) distance between component's peak activation and SOZ location (C – sLoreta, D – beamformer LCMV). Abbreviations used: A – amygdala, AG – angular gyrus, aI – anterior insula, IPL – inferior parietal lobule, PV – periventricular nodular heterotopia (in patient 9), MOFC – medial orbito-frontal cortex.